\documentclass[3p,times,procedia]{elsarticle}
\usepackage{nupha_ecrc}

\volume{00}

\firstpage{1}

\journalname{Nuclear Physics A}

\runauth{}

\jid{nupha}

\jnltitlelogo{Nuclear Physics A}

\usepackage{amssymb}

\usepackage{lineno}

\usepackage[figuresright]{rotating}

\usepackage{xspace}

\newcommand{\pt}{\mbox{$p_T$}\xspace}
\newcommand{\sqsntwo}{\mbox{$\sqrt{s_{NN}}=200$ GeV}\xspace}
\newcommand{\dca}{\mbox{$DCA$}\xspace}
\newcommand{\dcat}{\mbox{$DCA_T$}\xspace}
\newcommand{\dcal}{\mbox{$DCA_L$}\xspace}
\newcommand{\pp}{\mbox{$p$+$p$}\xspace}
\newcommand{\raa}{\mbox{$R_{AA}$}\xspace}

\begin{document}

\begin{frontmatter}



\dochead{}

\title{PHENIX Measurements of Single Electrons from Charm and Bottom Decays at Midrapidity in Au+Au Collisions}


\author{D. McGlinchey (for the PHENIX Collaboration)\footnote{For the full PHENIX Collaboration author list and acknowledgments,
see Appendix ``Collaboration'' of this volume}}

\address{University of Colorado, Boulder, Colorado 80309, USA}

\begin{abstract}

Heavy quarks are an ideal probe of the quark gluon plasma created in heavy ion collisions. They are produced in the initial hard scattering and therefore experience the full evolution of the medium. PHENIX has previously measured the modification of heavy quark production in Au+Au collisions at $\sqrt{s_{NN}}=200$ GeV via electrons from semileptonic decays, which indicated substantial modifications of the parent hadron momentum distribution. The PHENIX barrel silicon vertex detector (VTX), installed in 2011, allows for the separation of electrons from charm and bottom hadron decays through the use of displaced vertex measurements. These proceedings present the results of the completed analysis of the 2011 data set using the VTX.

\end{abstract}


\end{frontmatter}


\section{Introduction}
\label{sec:intro}

Heavy quarks, namely charm ($c$) and bottom ($b$), provide a probe of the coupling strength of the quark gluon plasma (QGP) formed in heavy ion collisions. Heavy quarks are produced dominantly in initial hard scatterings, while the lighter quarks ($u$, $d$, $s$) can be produced thermally in the medium. Heavy quarks, therefore, are clean probes which experience the full evolution of the QGP as they pass through it. Partons traversing the medium lose energy via both collisional and radiative energy loss. However, quarks experience a suppression of forward radiative energy loss, via the ``dead cone'' effect, which increases with the mass of the quark. This creates a natural expectation for the ordering of the energy loss ($\Delta E$) for quarks and gluons ($g$). Namely that $\Delta E_{g} > \Delta E_{u, d, s} > \Delta E_{c} > \Delta E_{b}$. By measuring the modification of heavy quark production in heavy ion collisions we can disentangle the contributions from collisional vs. radiative energy loss and more accurately constrain the medium coupling.

One channel for measuring the modification of heavy quark production is via electrons from semileptonic decays of heavy flavor mesons. Measurements by PHENIX at the Relativistic Heavy Ion Collider (RHIC) of electrons from semileptonic decays of combined charm and bottom hadrons as a function of transverse momentum (\pt) in Au+Au collisions at \sqsntwo indicate a strong suppression of the heavy quarks~\cite{Adare:2010de}. The PHENIX collaboration at RHIC installed a barrel silicon vertex detector (VTX) in 2011 with the goal of measuring electrons from separated charm and bottom hadron decays. An analysis of the data on Au+Au collisions at \sqsntwo collected in 2011, utilizing the VTX, coupled with the previous PHENIX measurement of the heavy flavor invariant yield~\cite{Adare:2010de} has been completed~\cite{Adare:2015hla}. The results were presented for the first time at Quark Matter 2015, and are discussed in these proceedings.

\section{Data Analysis and Unfolding}
\label{sec:analysis}

The analysis of electrons from separated charm and bottom hadron decays utilizes data on Au+Au collisions at \sqsntwo collected in 2011, as well as the previously published PHENIX heavy flavor electron invariant yield in Au+Au collisions~\cite{Adare:2010de} measured using data collected in 2004. The 2011 data adds precision tracking information from the VTX in the form of the distance of closest approach (\dca) of tracks to the collision vertex. The \dca is calculated separately in the transverse (\dcat) and longitudinal (\dcal) planes. The analysis is performed on \dcat, as it benefits from a greater resolution, $\sim60$ $\mu$m for tracks with $\pt>2$ GeV/$c$.

The \dcat distributions for electron candidates with $1.5<\pt\ [\mathrm{GeV}/c]<5.0$ are measured in five \pt bins over the quoted range. Background distributions from sources other than electrons from semileptonic charm and bottom hadron decays are determined and absolutely normalized using methods similar to the cocktail methods used in previous PHENIX electron analyses (see Ref.~\cite{Adare:2015hla} for details).

\begin{figure}
	\centering
	\includegraphics[width=0.35\textwidth]{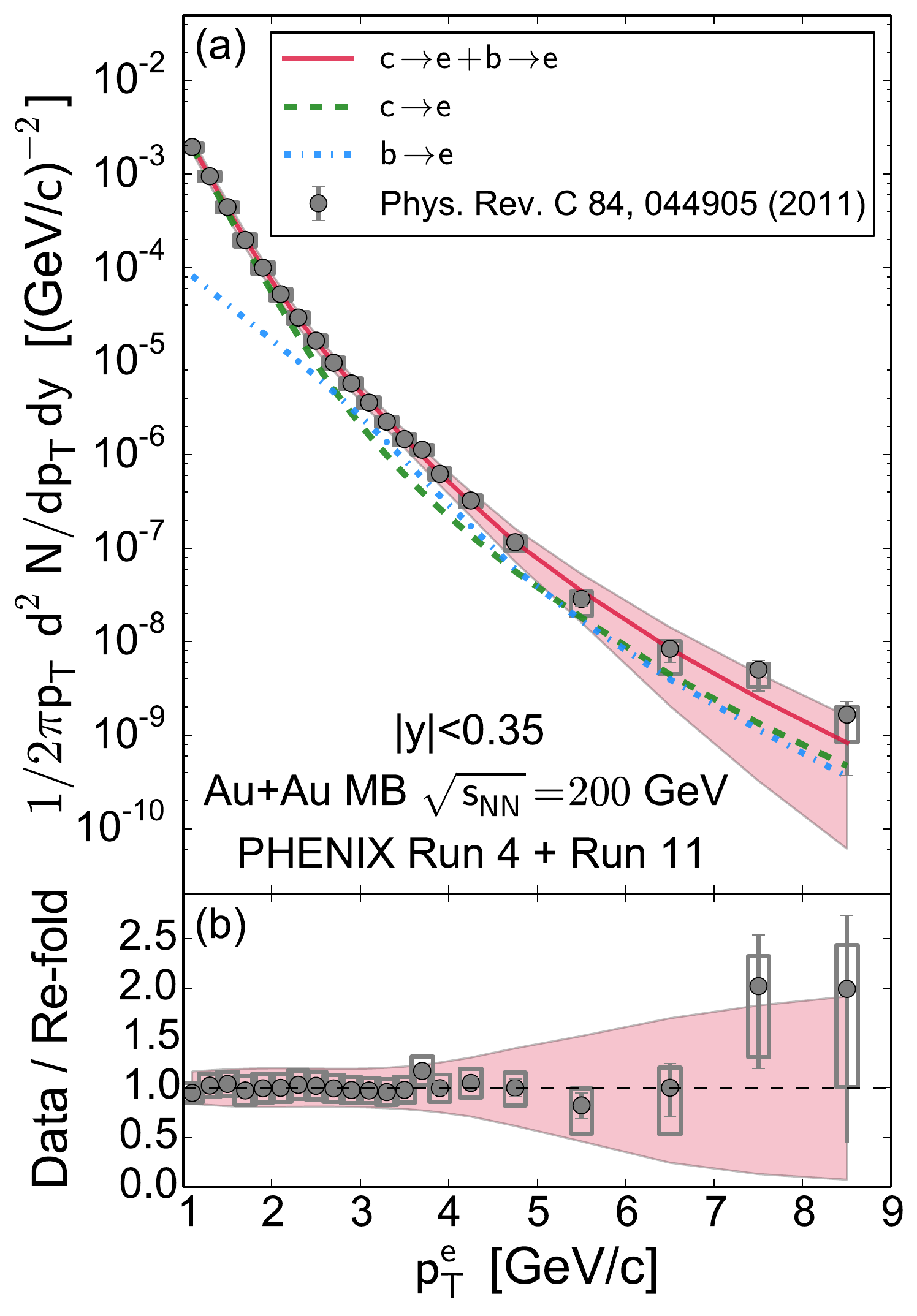}
	\includegraphics[width=0.53\textwidth, trim={0 4cm 0 4cm}, clip=true]{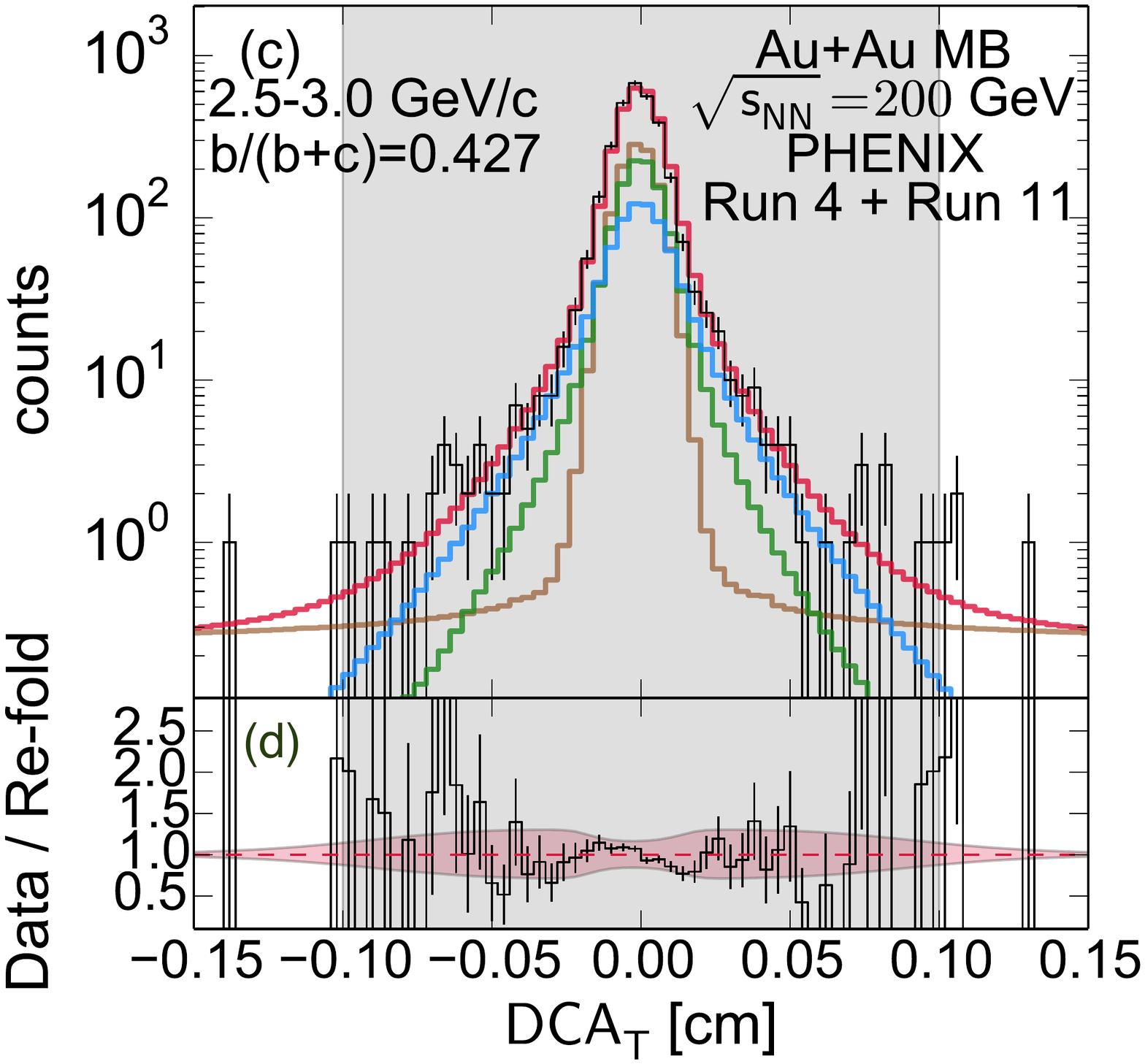}
	\caption{(a) The heavy flavor electron invariant yield as a function of \pt in Au+Au collisions at \sqsntwo~\cite{Adare:2010de} compared to the re-folded result. (b) The ratio of the data to the re-folded result as a function of \pt. (c) A comparison of the $DCA_T$ distribution for electron candidates in the range $2.5<\pt\ [\mathrm{GeV}/c]<3.0$ (black histogram) to the distributions for electrons from charm hadron decays (green line), bottom hadron decays (blue line), and background sources (brown line). (d) The ratio of the data to the sum of the background and unfolded electrons from charm and bottom. Comparisons of the $DCA_T$ distributions in other electron \pt bins can be found in Ref.~\cite{Adare:2015hla}.}
	\label{fig:refold}
\end{figure}

The electron \dcat distributions from the 2011 data set are fit simultaneously with the heavy flavor electron invariant yield using Bayesian inference techniques. This allows for a consistent description of heavy flavor electrons across both data sets, and as a function of \pt. The agreement between the measured data and the extracted distributions for electrons from separated charm and bottom hadron decays is shown in Fig.~\ref{fig:refold}. Good agreement is observed in both the heavy flavor electron invariant yield and the electron \dcat. 

\begin{figure}[!ht]
	\centering
	\includegraphics[width=0.7\textwidth]{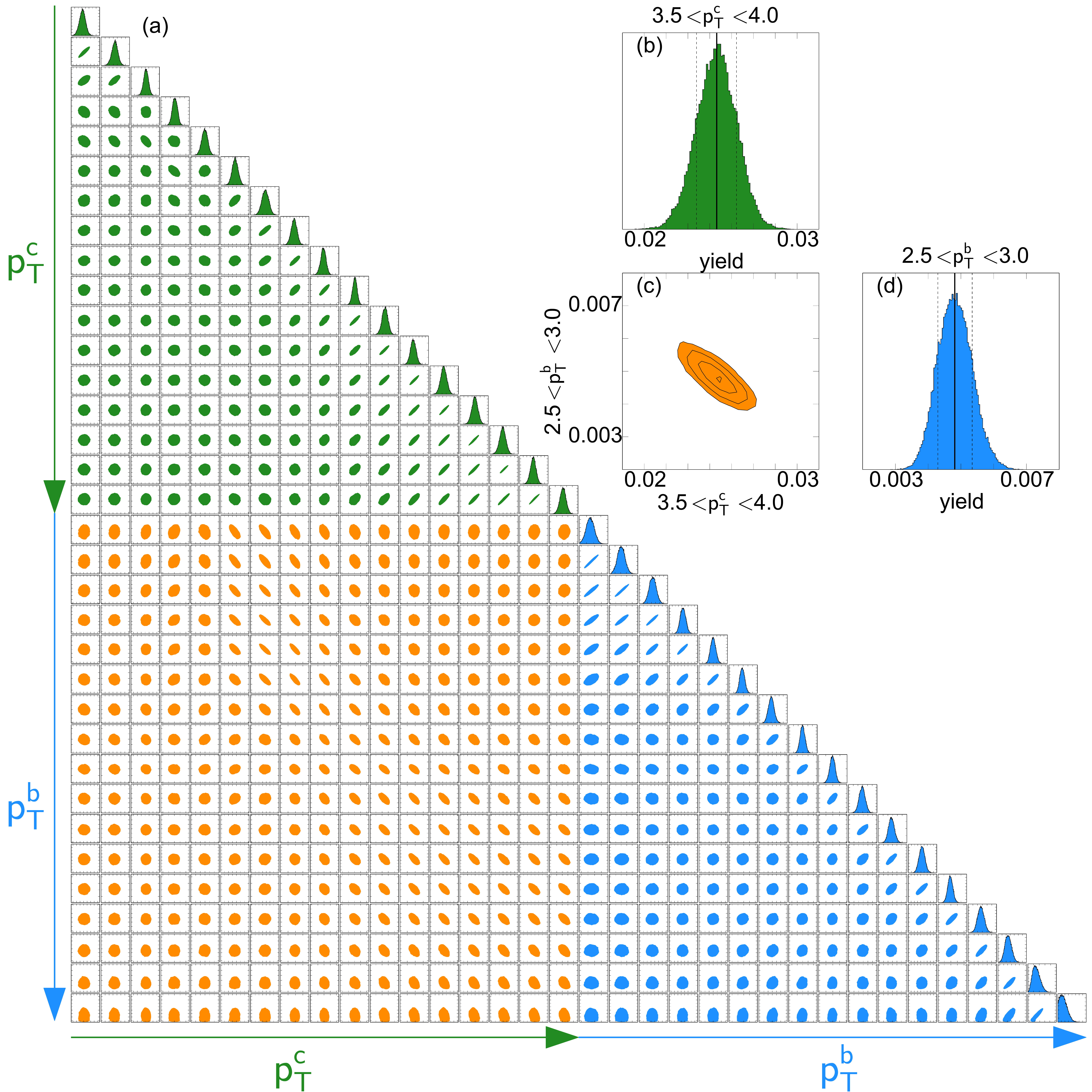}
	\caption{(a) The probability distributions for charm and bottom hadron yields in each \pt bin (diagonal) along with correlations between the yields (off-diagonal). The region of green plots (upper left quadrant) shows the charm hadron yields and the correlations between different charm hadron \pt bins. The region of blue plots (lower right quadrant) shows the bottom hadron yields and the correlations between different charm hadron \pt bins. The region of orange plots (lower left quadrant) shows the correlations between charm and bottom hadron yields in different \pt bins. The upper right shows an example correlation (c) between the charm hadron yield from $3.5<\pt\ [\mathrm{GeV}/c]<4.0$ (b) and the bottom hadron yield from $2.5<\pt\ [\mathrm{GeV}/c]>3.0$ (d)}
	\label{fig:posterior}
\end{figure}

The true parameters of the Bayesian inference techniques, sometimes referred to as unfolding, used in this analysis are the yields of charm and bottom hadrons in bins of \pt. A decay matrix, determined from {\sc pythia}, is used to propagate the hadron yields to electron space. One advantage of the unfolding used in this analysis is the extraction of the full probability distribution for the hadron yields in each \pt bin, as well as the correlations between the different parameters. This is shown in detail in Fig.~\ref{fig:posterior}. The diagonal plots show the probability distributions of charm or bottom hadrons in a given \pt bin, while the off-diagonal two dimensional plots show the correlations between them. For the correlation plots, a circular distribution indicates no correlation. An elliptical distribution with a positive slope indicates a positive correlation, and likewise, an elliptical distribution with a negative slope indicates an anti-correlation. Between the majority of hadron yields, there is little correlation. There is some positive correlation between adjacent bins in either charm or bottom hadron yields. This arises primarily from the regularization imposed, which enforces smooth hadron \pt distributions. There is also an interesting region of anti-correlation between charm and bottom hadron yields indicating that a trade-off between them gives an equally good fit to the electron data. A specific example is shown in Fig.~\ref{fig:posterior}(b)-(d).

\section{Results}
\label{sec:results}

\begin{figure}
	\centering
	\includegraphics[width=0.48\textwidth]{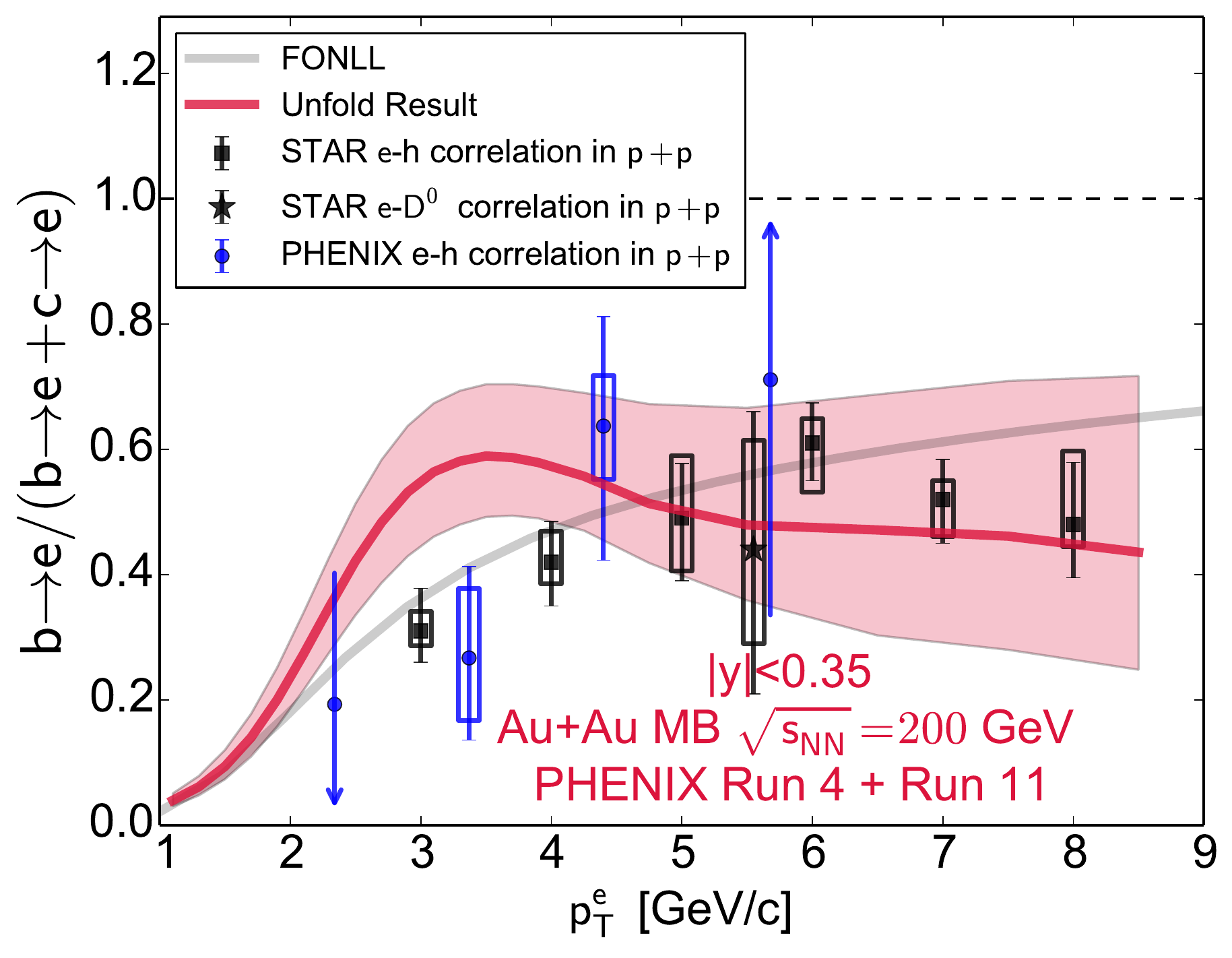}
	\includegraphics[width=0.45\textwidth]{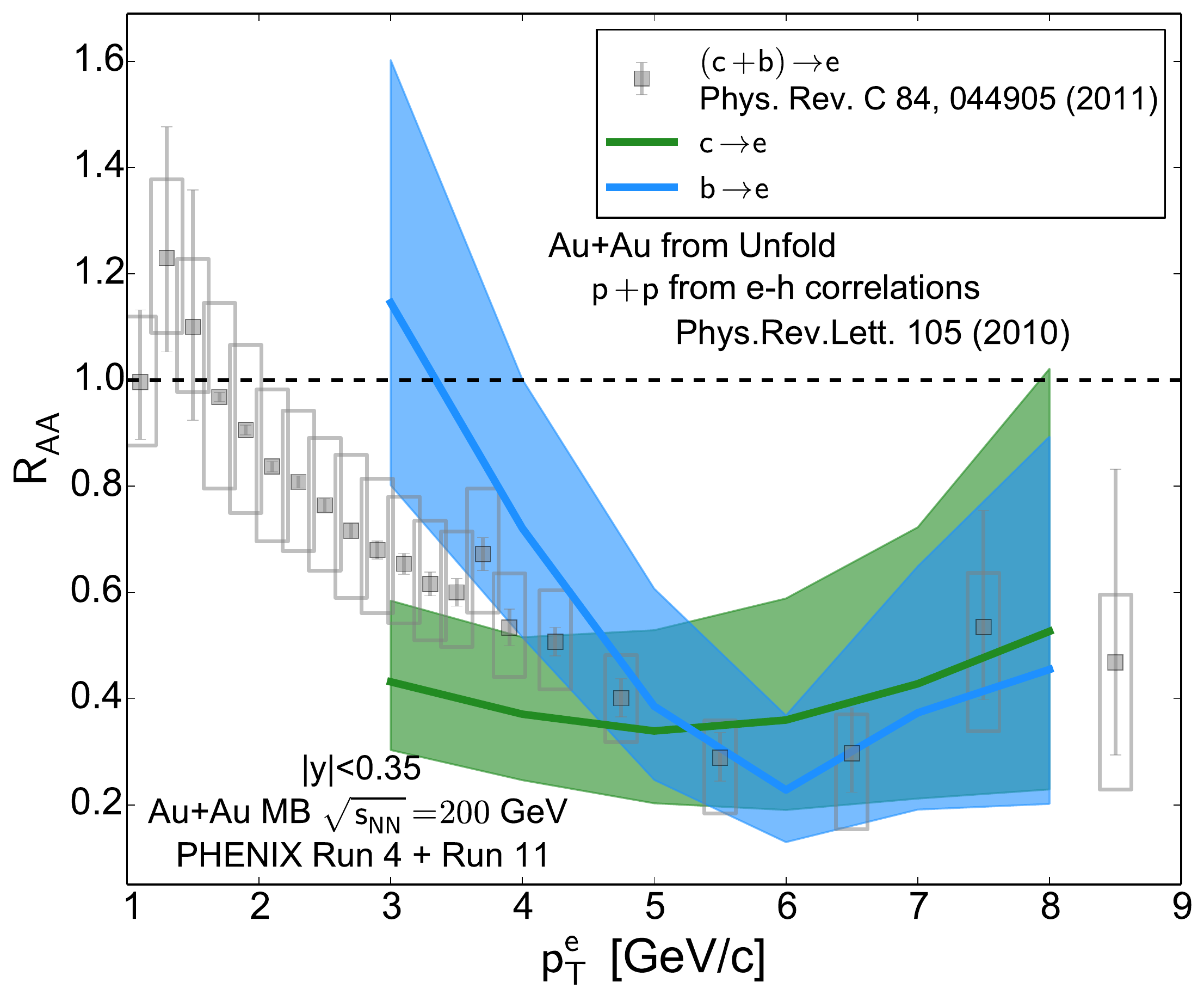}
	\caption{(Left) The unfolded bottom electron fraction as a function of electron \pt in Au+Au collisions at \sqsntwo~\cite{Adare:2015hla}. Also included are measurements of the bottom electron fraction in $p$+$p$ collisions at \sqsntwo from PHENIX~\cite{Adare:2009ic} and STAR~\cite{Aggarwal:2010xp}. (Right) The nuclear modification factor, $R_{AA}$, as a function of \pt for electrons from combined charm and bottom hadron decays~\cite{Adare:2010de}, as well as electrons from charm and bottom hadron decays separately~\cite{Adare:2015hla}.}
	\label{fig:hfraa}
\end{figure}

From the unfolded results for electrons from charm and bottom hadron decays, the bottom electron fraction is calculated. The results are shown in Fig.~\ref{fig:hfraa}(Left) compared to pQCD calculations of expected charm and bottom yields in \pp collisions from {\sc fonll}~\cite{Cacciari:2005rk}, therefore any deviation may indicate medium effects. A steep rise is observed for $\pt>2$ GeV/$c$, followed by a peak at $\pt\sim3.5$ GeV/$c$ and a small decrease for $\pt>4$ GeV/$c$. A larger bottom electron fraction is observed in Au+Au relative to {\sc fonll} at $\pt\sim3$ GeV/$c$. Also shown in Fig.~\ref{fig:hfraa}(Left) are measurements of the bottom electron fraction in \pp collisions at \sqsntwo~\cite{Adare:2009ic,Aggarwal:2010xp}, which are in good agreement with the {\sc fonll} calculations. 

The nuclear modification factor, \raa, for electrons from charm, and separately bottom, hadron decays can be calculated from the total heavy flavor \raa, and the bottom electron fraction in Au+Au and \pp (see Ref.~\cite{Adare:2015hla} for details). The results are shown in Fig.~\ref{fig:hfraa}(Right). A larger suppression of electrons from charm hadron decays compared to bottom hadron decays is found for $\pt<4$ GeV/$c$. However, the relatively large uncertainties make detailed comparisons difficult at this time.

\section{Summary}
\label{sec:summary}

Using the VTX detector installed in 2011, PHENIX has measured the yield of electrons from separated charm and bottom hadron decays~\cite{Adare:2015hla}. A larger suppression of electrons from charm hadron decays compared to bottom hadron decays is found for $\pt<4$ GeV/$c$, with a similar suppression observed for $\pt>4$ GeV/$c$. These results were extracted using Bayesian inference techniques which allowed for the extraction of the \pt distributions of the parent hadrons. This includes not only mean values with uncertainties, but full probability distributions and correlations between the different \pt bins. 

This is the first measurement that bottom quarks are significantly suppressed in Au+Au collisions at \sqsntwo. However, the uncertainties on the current measurement preclude a detailed understanding of the ordering of the charm and bottom suppression for $\pt>4$ GeV/$c$. However, PHENIX collected data on Au+Au collisions at \sqsntwo in 2014 which represents an order of magnitude increase in the electron statistics. A similar analysis of the 2014 data set, coupled with \pp data collected in 2015, should provide a definitive measurement of the separated heavy flavor electron suppression.





\bibliographystyle{elsarticle-num}
\bibliography{McGlinchey_QM2015_proceedings}







\end{document}